\documentstyle[aps,pra,epsf,preprint]{revtex}

\begin{document}

\title{A prognosis oriented microscopic stock market model}

\author{Christian Busshaus$^1$ and Heiko Rieger$^{1,2}$}
\address{
$^1$ Institut f\"ur Theoretische Physik, Universit\"at zu K\"oln, 
50923 K\"oln, Germany\\
$^2$ NIC c/o Forschungszentrum J\"ulich, 52425 J\"ulich, Germany
}

\date{February 25, 1999}

\maketitle

\begin{abstract}
  We present a new microscopic stochastic model for an ensemble of
  interacting investors that buy and sell stocks in discrete time steps
  via limit orders based on individual forecasts about the price of
  the stock. These orders determine the supply and demand fixing after
  each round (time step) the new price of the stock according to which
  the limited buy and sell orders are then executed and new forecasts
  are made. We show via numerical simulation of this model that the
  distribution of price differences obeys an exponentially truncated
  Levy-distribution with a self similarity exponent
  $\mu\approx5$.\\ \end{abstract}


truncated Levy distribution.

\newcommand{\bc}{\begin{center}}
\newcommand{\ec}{\end{center}}
\newcommand{\bi}{\begin{itemize}}
\newcommand{\ei}{\end{itemize}}
\newcommand{\be}{\begin{equation}}
\newcommand{\ee}{\end{equation}}
\newcommand{\beqn}{\begin{eqnarray}}
\newcommand{\eeqn}{\end{eqnarray}}

\section{Introduction}

In the last years a number of microscopic models for price
fluctuations have been developed by physicists
\cite{Levy_Levy_Solomon,Cont_Bouchaud,Bak_Paczuski_Shubik,Caldarelli_Marsili_Zhang,Chowdhury_Stauffer,Oliveira_Stauffer} and economists \cite{kim,lux}.
The purpose of these models is, in our view, not to make specific
predictions about the future developments of the stock market (for
instance with the intention to make a fortune) but to reproduce the
universal statistical properties of liquid markets.

Some of these properties are an exponentially truncated
Levy-distribution for the price differences on short time scales
(significantly less than one month) and a linear autocorrelation
function of the prices which decays to zero within a few minutes
\cite{Bouchaud,Mantegna_Stanley,Meyer_Stanley,Matacz,Cont_Potters_Bouchaud}.

We present a new microscopic model with interacting investors in the
spirit of \cite {lux,Cont_Bouchaud,farmer} that speculate on price
changes that are produced by themselves. The main features of the
model are individual forecasts (or prognoses) for the stock price in
the future, a very simple trading strategy to gain profit, {\it
  limited orders} for buying and selling stocks \cite{kim} and various
versions of interaction among the investors during the stage of
forecasting the future price of a stock.

The paper is organized as follows: In section 2 we define our model,
in section 3 we present the results of numerical simulations of this
model including specific examples of the price fluctuations using
different interactions among the investors, the autocorrelation
function of the price differences and most importantly their
distribution, which turn out to be (exponentially) truncated Levy
distributions. Section 4 summarizes our findings and provides an
outlook for further refinements of the model.

\section{The model}

The system consists of one single stock with actual price $K(t)$ and
$N$ investors labeled by an index $i=1,\ldots,N$.  In the
most simplified version of the model the investors have identical
features and are described at each time step by three variables: 
\bi
\item[$P_i(t)$] The personal prognosis of investor $i$ at time $t$ about
  the price of the stock at time $t+1$.
\item[$C_i(t)$] The cash capital (real variable) of investor $i$ at time $t$.
\item[$S_i(t)$] The number of shares (integer variable) of investor $i$ at time $t$.
\ei
The system at time $t=0$ is initialized with some appropriately
generated initial values for $P_i(t=0)$, $C_i(t=0)$ and $S_i(t=0)$,
plus a particular price for the stock.

The dynamics of the system evolves in discrete time steps
$t=1,2,3,\ldots$ and is defined as follows. Suppose time step $t$ has
been finished, i.e.\ the variables $K(t)$, $P_i(t)$, $C_i(t)$ and
$S_i(t)$ are known. Then the following consecutive procedures are
executed.

\noindent
\underline{\bf Make Prognosis}\\
Each investor sets up a new personal prognosis via
\be\label{eq_prognosis}
P_i(t+1)=(xP_i(t)+(1-x)K(t))\cdot e^{r_i},
\ee
where $x\in[0,1]$ is a model dependent weighting factor (for the
investor's old prognosis and the price of the stock) and $r_i$ are
independent identically distributed random variables of mean zero and
variance $\sigma$ that mimic a (supposedly) stochastic component in
the individual prognosis (external influence, greed, fear, sentiments
$\cdots$, see also \cite{kim}).

\noindent
\underline{\bf Make Orders}\\
Each investor gives his limit order on the basis of his old and his
new prognosis:\\
$P_i(t+1)-P_i(t)>0$:\\ 
investor $i$ puts a buy-order limited by $P_i(t)$, which means that he
wants to transform all cash $C_i(t)$ into ${\rm int}[C_i(t)/P_i(t)]$
shares {\bf if} $K(t+1)\leq P_i(t)$.\\ 
$P_i(t+1)-P_i(t)<0$:\\ 
investor $i$ puts a sell-order limited by $P_i(t)$, which means that he
wants to transform all stocks into $S_i(t)\cdot K(t+1)$ cash {\bf if}
$K(t+1)\geq P_i(t)$.\\ 
Now let $i_1,i_2,\ldots,i_{N_A}$ be the investors that have put a
sell-order and their limits are $P_{i_1}(t)\le P_{i_2}(t)\le\cdots\le
P_{i_{N_A}}(t)$, and let $j_1,j_2,\ldots,j_{N_B}$ be the investors that
have put a buy-order and their limits are $P_{j_1}(t)\ge
P_{j_2}(t)\ge\cdots\ge P_{j_{N_B}}(t)$. 

\noindent
\underline{\bf Calculate new price}\\
Define the supply and demand functions $A(K)$ and $B(K)$,
respectively, via
\beqn
A(K) & = & \sum_{a=1}^{N_A} S_{i_a} \cdot \theta(K-P_{i_a}(t))
\nonumber\\
B(K) & = & \sum_{b=1}^{N_A} \Delta S_{j_b} \cdot
[1-\theta(K-P_{j_b}(t))]
\eeqn
with $\Delta S_{j_b}={\rm int}[C_{j_b}(t)/P_{j_b}(t)]$ the number of
shares demanded by investor $j_b$, and $\theta(x)=1$ for $x\ge0$ and
$\theta(x)=0$ for $x<0$. Then the total turnover at price $K$ would be
\be
Z(K)={\rm min}\,\{A(K),B(K)\}
\ee
and the new price is determined is such a way that $Z(K)$ is
maximized. Since $Z(K)$ is a piece-wise constant function it is maximal
in a whole interval, say $K\in[P_{i_{\rm max}},P_{j_{\rm max}}]$ for
some $i_{\rm max}\in\{i_1,\ldots,i_{N_A}\}$ and $j_{\rm
max}\in\{j_1,\ldots,i_{N_B}\}$. Then we define the new price to be
the weighted mean
\be
K(t+1)=\frac{P_{i_{\rm max}}\cdot A(P_{i_{\rm max}}) + 
             P_{j_{\rm max}}\cdot B(P_{j_{\rm max}})}
            {A(P_{i_{\rm max}}) + B(P_{j_{\rm max}})}\;.
\ee
Note that the weight by the total supply and demand takes care of the
price being slightly higher (lower) than the arithmetic mean
$(P_{i_{\rm max}}+P_{j_{\rm max}})/2$ if the supply is smaller
(larger) than the demand.

\noindent
\underline{\bf Execute orders}\\
Finally the sell-orders of the investors $i_1,\ldots,i_{\rm max}$ and the
buy-orders of the investors $j_1,\ldots,j_{\rm max}$ are executed at the new price $K(t+1)$, i.e. the buyers $j_1,\ldots,j_{\rm max}$ update
\beqn
S_{j_b}(t+1) & = & S_{j_b}(t)+{\rm int}[C_{j_b}(t)/P_{j_b}(t)]
\nonumber\\
C_{j_b}(t+1) & = & C_{j_b}(t)-K(t+1)\cdot(S_{j_b}(t+1)-S_{j_b}(t))
\eeqn
and the investors $i_1,\ldots,i_{\rm max}$ sell all their shares at
price $K(t+1)$: 
\beqn
S_{i_a}(t+1) & = & 0
\nonumber\\
C_{i_a}(t+1) & = & C_{i_a}(t)+S_{i_a}(t)\cdot K(t+1)
\eeqn
\noindent
If $A(P_{i_{\rm max}})<B(P_{j_{\rm max}})$ then investor $j_{\rm max}$
cannot buy ${\rm int}[C_{j_{\rm max}(t)}/P_{j_{\rm max}}(t)]$ 
but only the remaining shares, whereas in the case 
$A(P_{i_{\rm max}})>B(P_{j_{\rm max}})$ investor $i_{\rm max}$ cannot 
sell all his shares.
\noindent 
The orders of the investors $i_{{\rm max}+1},\ldots,i_{N_A}$ and 
$j_{{\rm max}+1},\ldots,j_{N_B}$ cannot be executed due to their limits.

The execution of orders completes one round, measurements of
observables can be made and then the next time step will be processed.

A huge variety of interaction among the investors can be modeled, here
we restrict ourselves to three different versions taking place at the
level of the individual prognosis genesis:

\bi
\item[I$_1$:] 
  Each investor $i$ knows the prognoses 
  $P_{i_1}(t),\ldots,P_{i_m}(t)$ of $m$ randomly selected (once at the 
  beginning of the simulation) neighbors. When making an order,
  he modifies his strategy and puts in the case
  \be
  P_i(t+1)-[g_i(t) P_i(t)+\sum_{n=1}^{m}g_{i_n}(t) P_{i_n}(t)]<(>)0
  \ee
  \noindent
  a buy (sell) order limited still by his own prognosis $P_i(t)$. 
  We choose the weights $g_i(t)=1/2$ and $g_{i_n}(t)=1/2m$ for
  $n=1,\ldots,m$.
  \item[I$_2$:] 
  In addition to interaction I$_1$ investor $i$ changes the weights
  $g$ after the calculation of the new price $K(t+1)$ according to
  the success of the prognoses:
  \beqn
  g_{i_-}(t+1)=g_{i_-}(t)-\Delta g
  \nonumber\\
  g_{i_+}(t+1)=g_{i_+}(t)+\Delta g
  \eeqn
  \noindent
  where fro each investro $i$ the index $i_-$ ($i_+$) denotes the 
  investor from the set $\{i,i_1,\ldots,i_m\}$ with the worst (best) 
  prognosis, i.e.:
  \beqn
  i_-\in\{i,i_1,\ldots,i_m\}\qquad{\rm such\quad that}\qquad
  {\rm abs}[P_{i_-}(t)-K(t+1)]\qquad{\rm is\quad maximal}
  \nonumber\\
  i_+\in\{i,i_1,\ldots,i_m\}\qquad{\rm such\quad that}\qquad
  {\rm abs}[P_{i_+}(t)-K(t+1)]\qquad{\rm is\quad minimal}
  \eeqn
  \noindent
  The weight $g_i$ is forced to be positive, because an investor 
  should believe in his own prognosis $P_i(t)$.
\item[I$_3$:] 
  In addition to interaction I$_2$ neighbors with weights $g_{i_-}(t+1)<0$
  are replaced by randomly selected new neighbors.
\ei

\section{Results}

In this section we present the results of numerical simulations of the
model described above. In what follows we consider a system with 1000
investors and build ensemble averages over 10000 independent samples
(i.e.\ simulations) of the system. We checked that the results we are
going to present below do not depend on the system size (the number of
traders). When changing the system size, i.e. the number $N$ of
investors, the statistical properties of the price differences do not
change qualitatively. Increasing $N$ only decreases the average
volatility (variance of the price changes).

For concreteness we have chosen the following parameters: the initial
price of the stock is $K_0=100$ (arbitrary units, \cite{kim}), Each trader has
initially $C_i(t=0)=50000$ units of cash and $S_i(t=0)=500$ stocks
(thus the total capital of each trader is initially 100000 units). The
standard deviation of the Gaussian random variable $z$ is
$\sigma=0.01$ (with mean zero). We performed the simulations over 1000
time steps which is roughly 10 time longer than the transient time of
the process for these parameters. In other words, we are looking at
its stationary properties.

First we should note that in the deterministic case $\sigma=0$ no
trade would take place \cite{Levy_Levy_Solomon}, hence the stochastic
component in the individual forecasts is essential for any interesting
time evolution of the stock market price.

We focus on the time dependence of the price $K(t)$, the price change
$\Delta_T(t)=K_{t+T}-K_T$ in an interval $T$, their time dependent
autocorrelation
\be
C_T(\tau)=\frac{
\langle\Delta_T(t+\tau)\Delta_T(t)\rangle
\langle\Delta_T(t+\tau)\rangle\langle\Delta_T(t)\rangle}
{\langle(\Delta_T(t))^2\rangle-\langle\Delta_T(t)\rangle^2}
\ee
and their probability distribution $P(\Delta_T(t))$. The statistical
properties of the price changes produced by our model depend very
sensitively on the parameter $x$ in equation~(\ref{eq_prognosis}). In
particular for the case $x=1$ it turns out that the total turnover
decays like $t^{-1/2}$ in the interaction-free case, which implies
that after a long enough time no investor will buy or sell anything
anymore. However, only an infinitesimal deviation from $x=1$ leads to
a saturation of the total turnover at some finite value and trading
will never cease.

In Fig.1--4 we present the results of the interaction-less case with
$x=1$ (Fig. 1) and $x=0$ and contrast it with the results of the model
with interactions $I_1$, also for $x=1$ (Fig. 3) and $x=0$ (Fig. 4).

For $x=0$ investor $i$ does not look at his old prognosis but only at
the actual stock price when making a new prognosis. In this case the
distribution of the price can be fitted very well by a Gaussian
distribution irrespective of the version of interaction or no
interaction. The self similarity exponent $1/\mu\approx 0.5$ agrees
with the scaling behavior of a Gaussian distribution.  The
autocorrelation function of the price differences decays alternating
to zero within a few time steps.

In the opposite case $x=1$ investor $i$ makes his new prognosis
$P_i(t+1)$ based on his own old one and never looks at the current
stock price. Now we can show that the distribution of the price
differences decays exponentially in its asymptotic, but the self
similarity exponent $1/\mu\approx 0.2$ is too small to agree with a
Levy stable distribution. The autocorrelation function of the price
differences decays very quickly, so that there are significant linear
anti-correlations only between consecutive differences.

\bc
\begin{tabular}{|c||c|c|c|c|}
\hline
$1/\mu$ & I$_0$ & I$_1$ & I$_2$ & I$_3$ \\
\hline
\hline
$x=0$ & $0.442$ & $0.466$ & $0.472$ & $0.472$ \\
$x=1$ & $0.228$ & $0.212$ & $0.185$ & $0.185$ \\
\hline
\end{tabular}
\ec

The selfsimilarity exponent has been determined via the scaling
relation $P(\Delta_T=0)\sim T^{-1/\mu}$ and a linear fit to the data
of $P(\Delta_T=0)$ versus $T$ in a log-log plot. These least square
fits yield the relative errors for our estimates of the self
similarity exponent $1/\mu$ in the table above, which lay between
$0.1\%$ and $0.3\%$.

\section{Summary and outlook}

We presented a new microscopic model for liquid markets
that produces an exponentially truncated Levy-distribution
with a self similarity exponent $1/\mu\approx 0.2$ for the
price differences on short time scales. Studying the distribution
on longer time scales we find that it converges to a Gaussian
distribution. The autocorrelation function of the price changes
decays to zero within a few time steps. The statistical properties
of our prognosis oriented model depend very sensitively on the rules
how the investors make their prognoses.

There are many possible variations of our model that could be studied.
It is plausible that a heterogeneous system of traders leads to
stronger price fluctuations and thus a smaller value for the self
similarity exponent $\mu$ (which appears to be $1/\mu\approx 0.7$ for
real stock price fluctuations \cite{Mantegna_Stanley}). The starting
wealth could be distributed with a potential law (comparable with the
cluster size in the Cont-Bouchaud model). Or the investors could have
different rules for making prognoses and following trading strategies.
Another possible variation is to implement a threshold in the simple
strategy in order to simulate risk aversion (the value of the
threshold could depend on the actual volatility).

Unfortunately, forecasts for real stock markets cannot be made
with our model, because it is a stochastic model. We see possible
applications for this model in the pricing and the risk measurement
of complex financial derivatives.

{\bf Acknowledgment} We thank D. Stauffer for helpful discussions.
H.\ R.'s work was supported by the Deutsche Forschungsgemeinschaft
(DFG).

\vfill
\eject

\begin{figure}
\epsfxsize=8cm\epsfbox{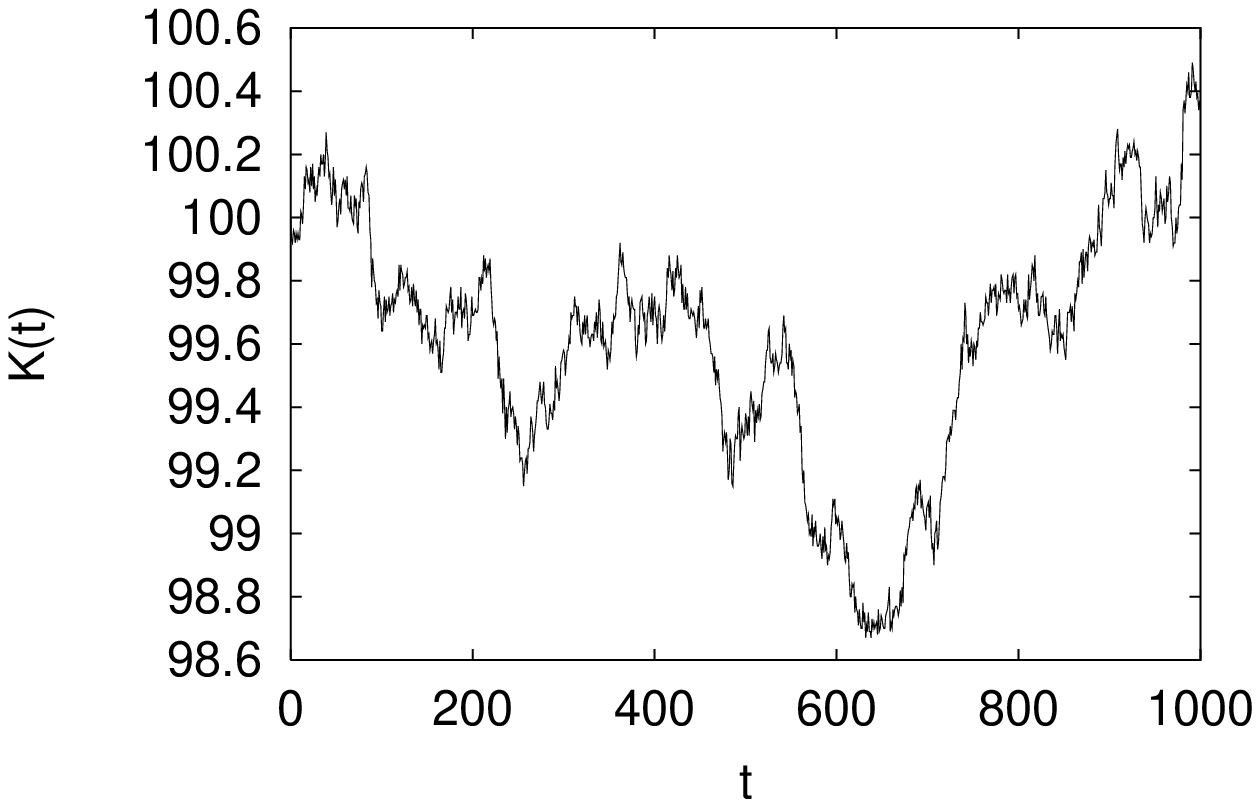}
\epsfxsize=8cm\epsfbox{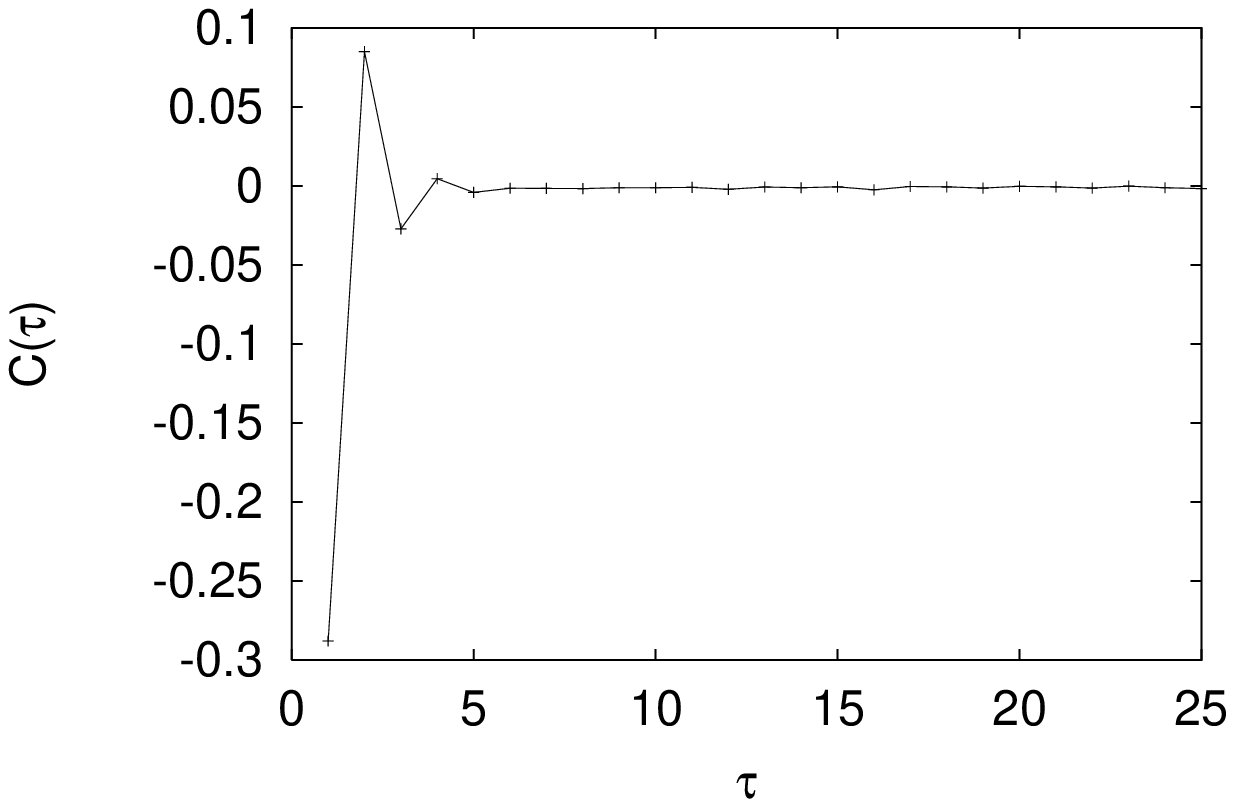}
\epsfxsize=8cm\epsfbox{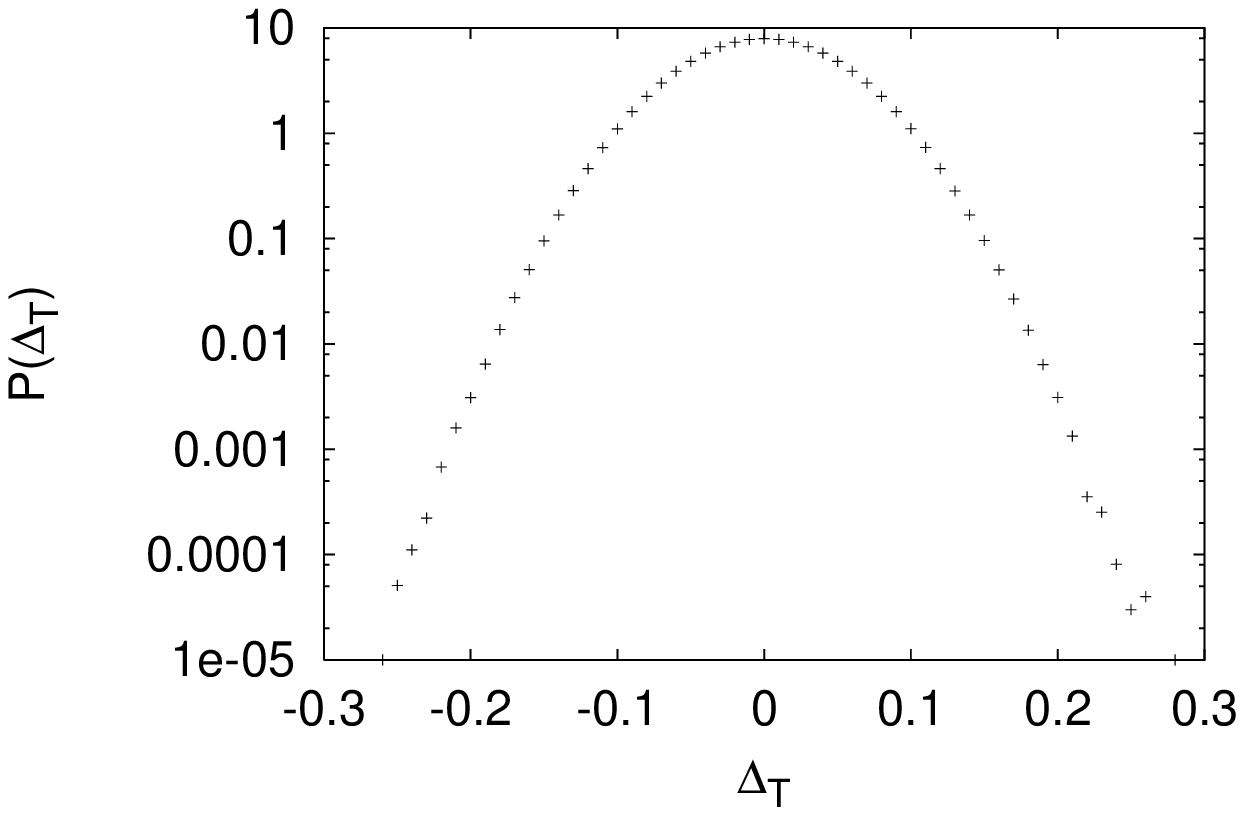}
\caption{Results of numerical simulations for the model {\it without}
interactions $I_0$ and $x=0$ (i.e.\ investors look only at their old
prognosis $P_i(t)$). Shown are the price fluctuations for one sample
(top), the autocorrelation function $C_T(\tau)$ for $T=1$ (middle) and
the probability distribution $P(\Delta_T)$ of the price differences
for $T=1$.
\label{fig1}
}
\end{figure}
\vfill
\eject

\begin{figure}
\epsfxsize=8cm\epsfbox{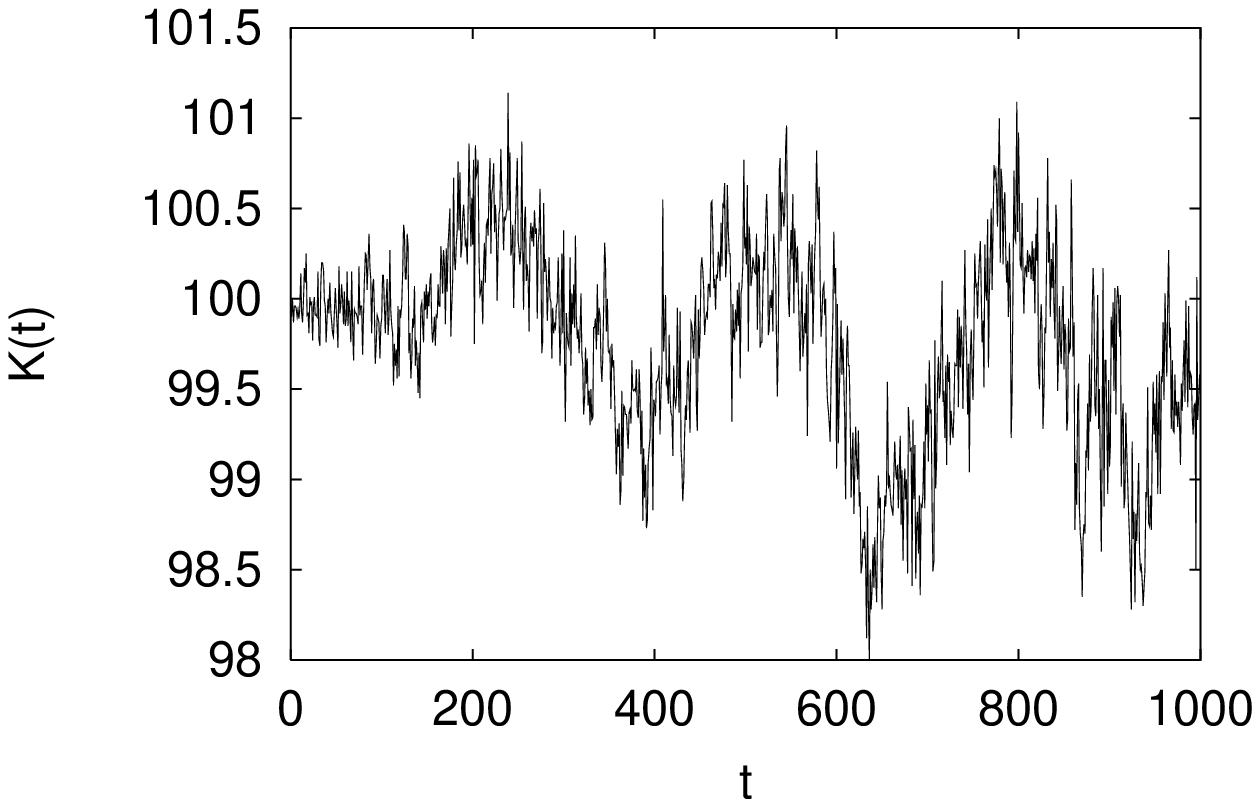}
\epsfxsize=8cm\epsfbox{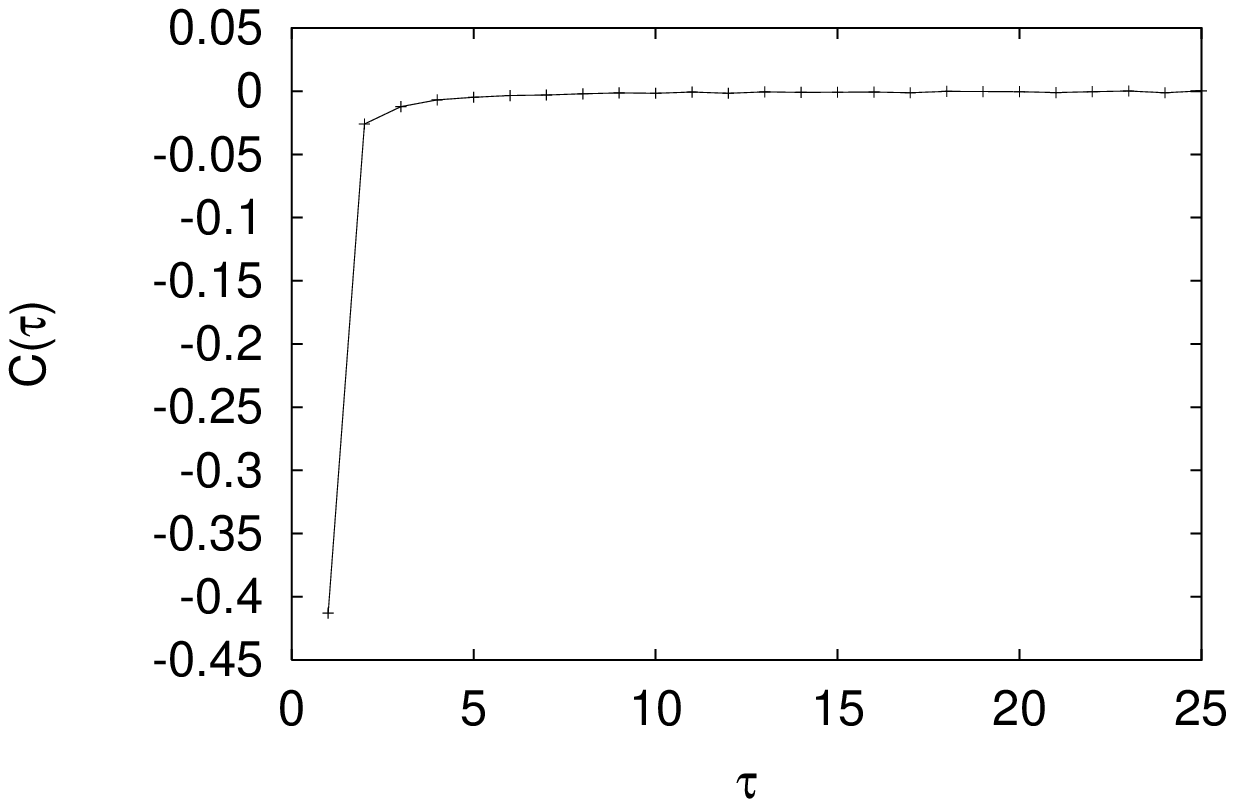}
\epsfxsize=8cm\epsfbox{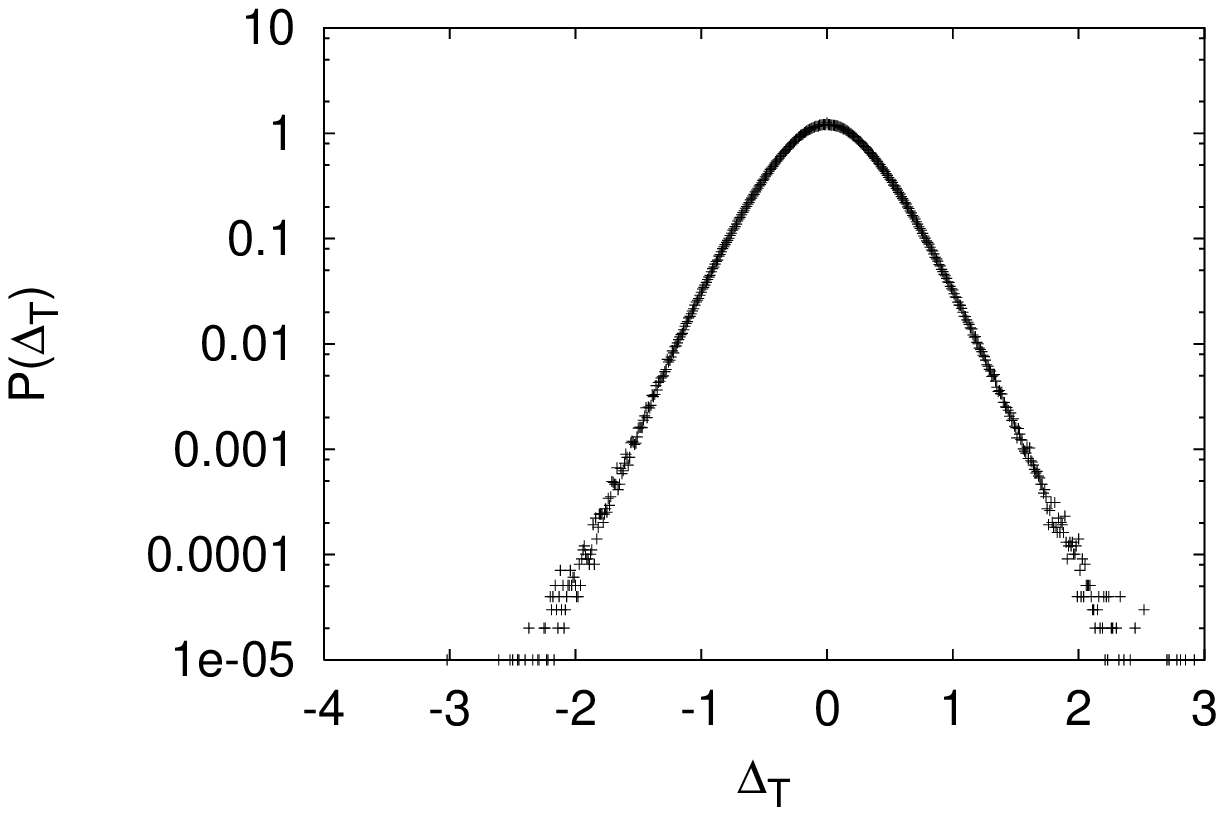}
\caption{The same as Fig. 1, however with $x=1$ (i.e. investors look
only at the old price $K(t)$).
\label{fig2} 
}
\end{figure}
\vfill
\eject

\begin{figure}
\epsfxsize=8cm\epsfbox{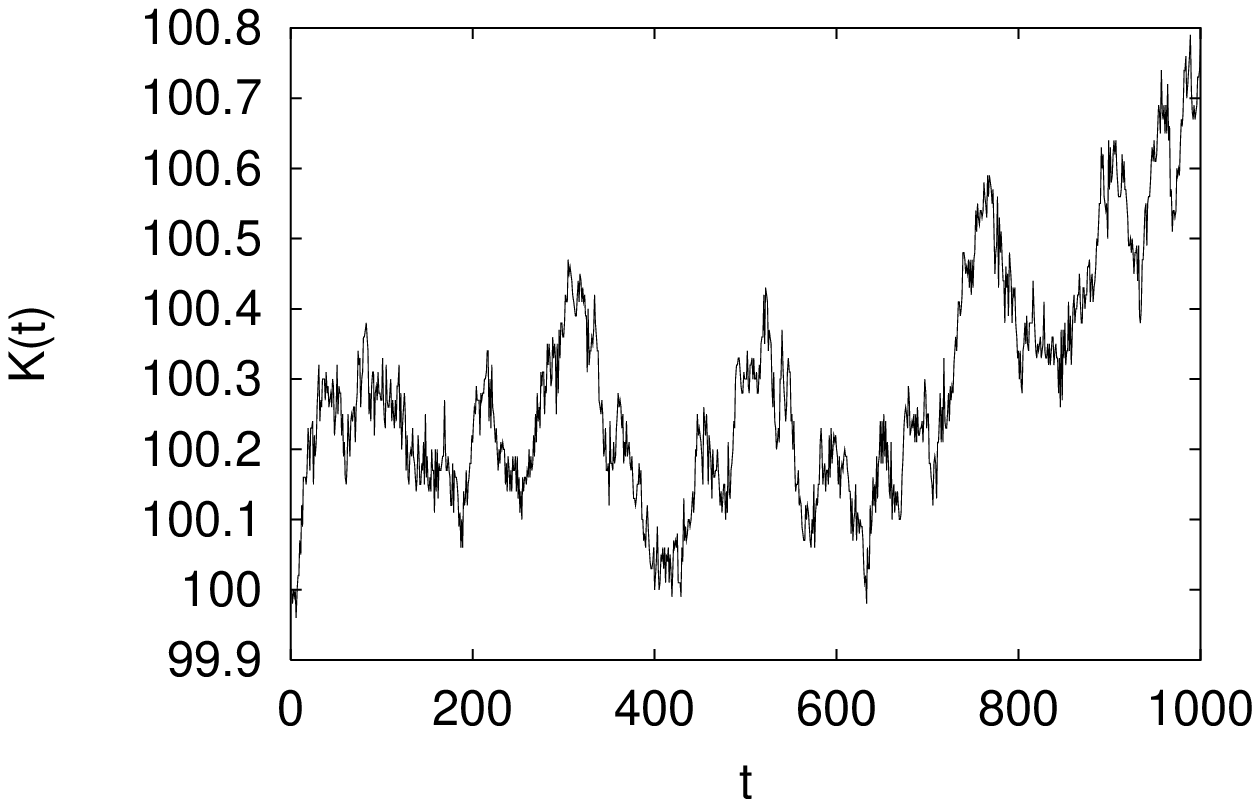}
\epsfxsize=8cm\epsfbox{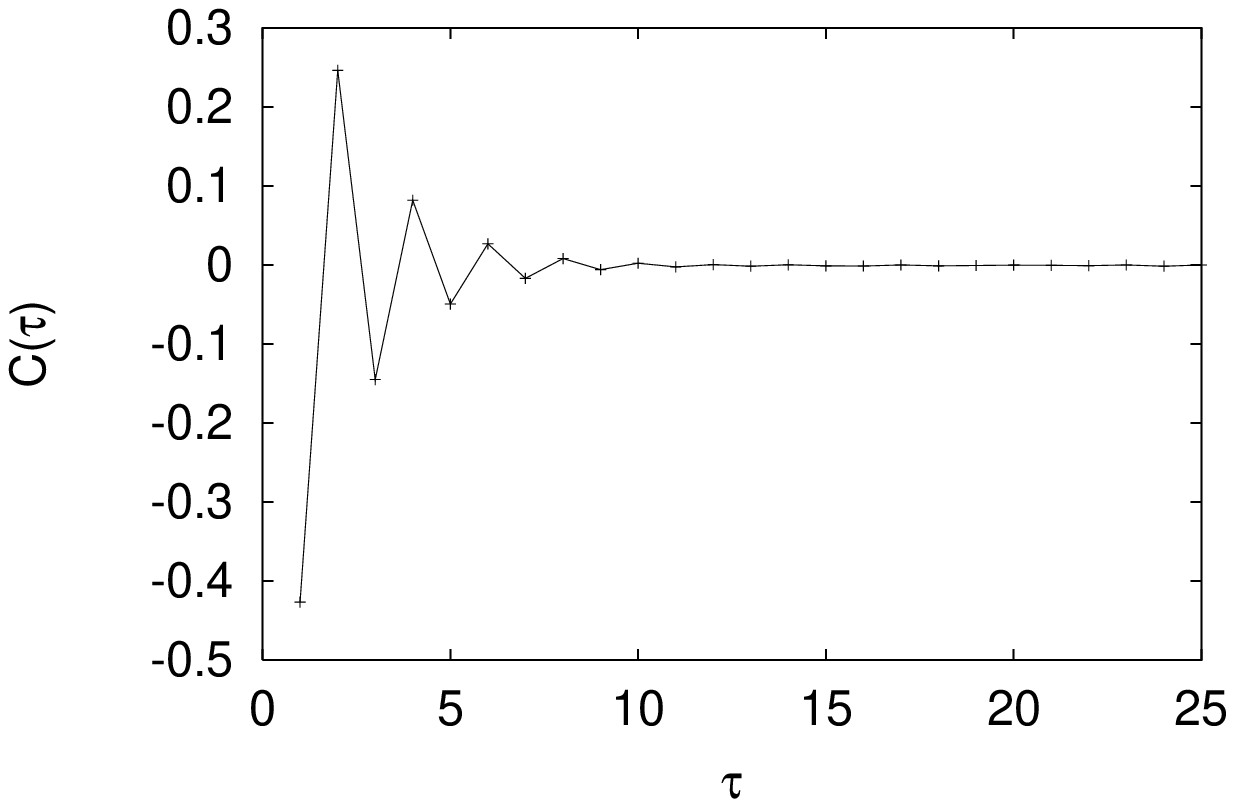}
\epsfxsize=8cm\epsfbox{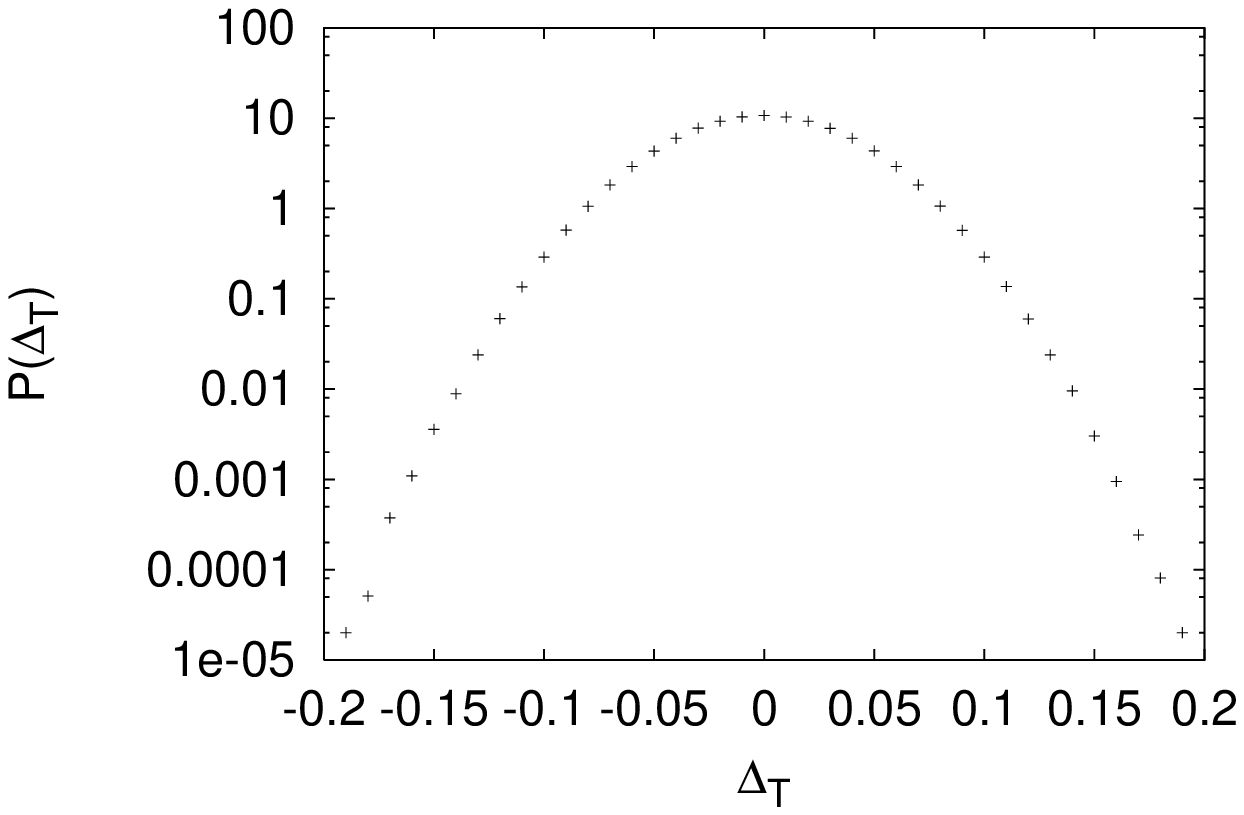}
\caption{The same as Fig. 1, however with interactions $I_1$ (see
text) and $x=0$ (i.e. investors look only at their old prognosis
$P_i(t)$).
\label{fig3}
}
\end{figure}
\vfill
\eject

\begin{figure}
\epsfxsize=8cm\epsfbox{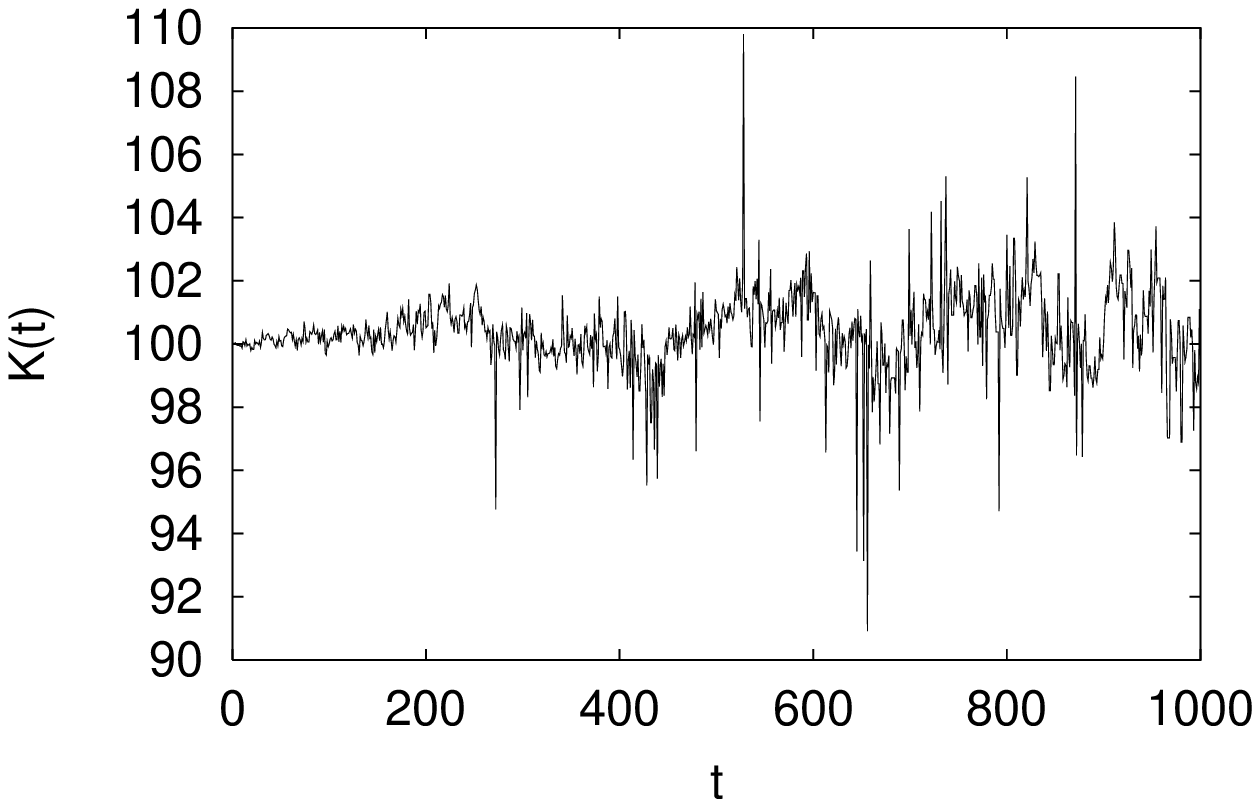}
\epsfxsize=8cm\epsfbox{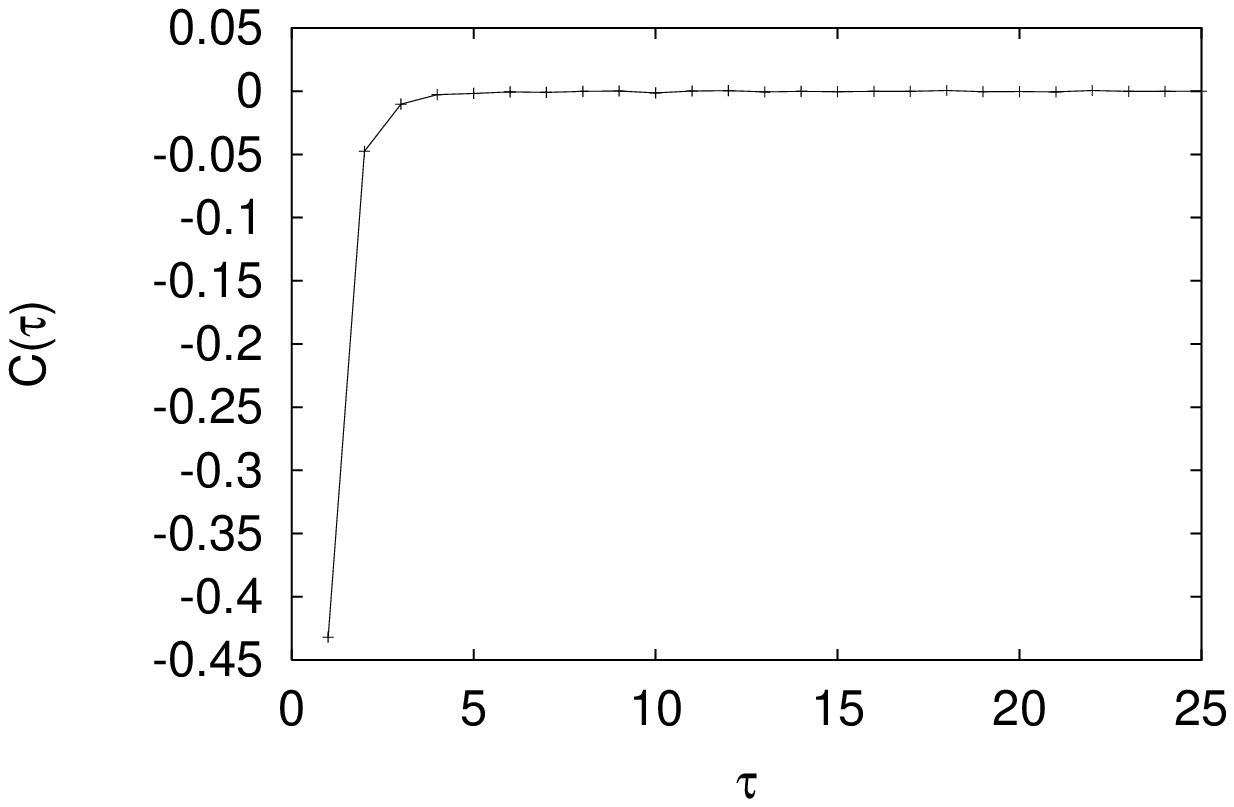}
\epsfxsize=8cm\epsfbox{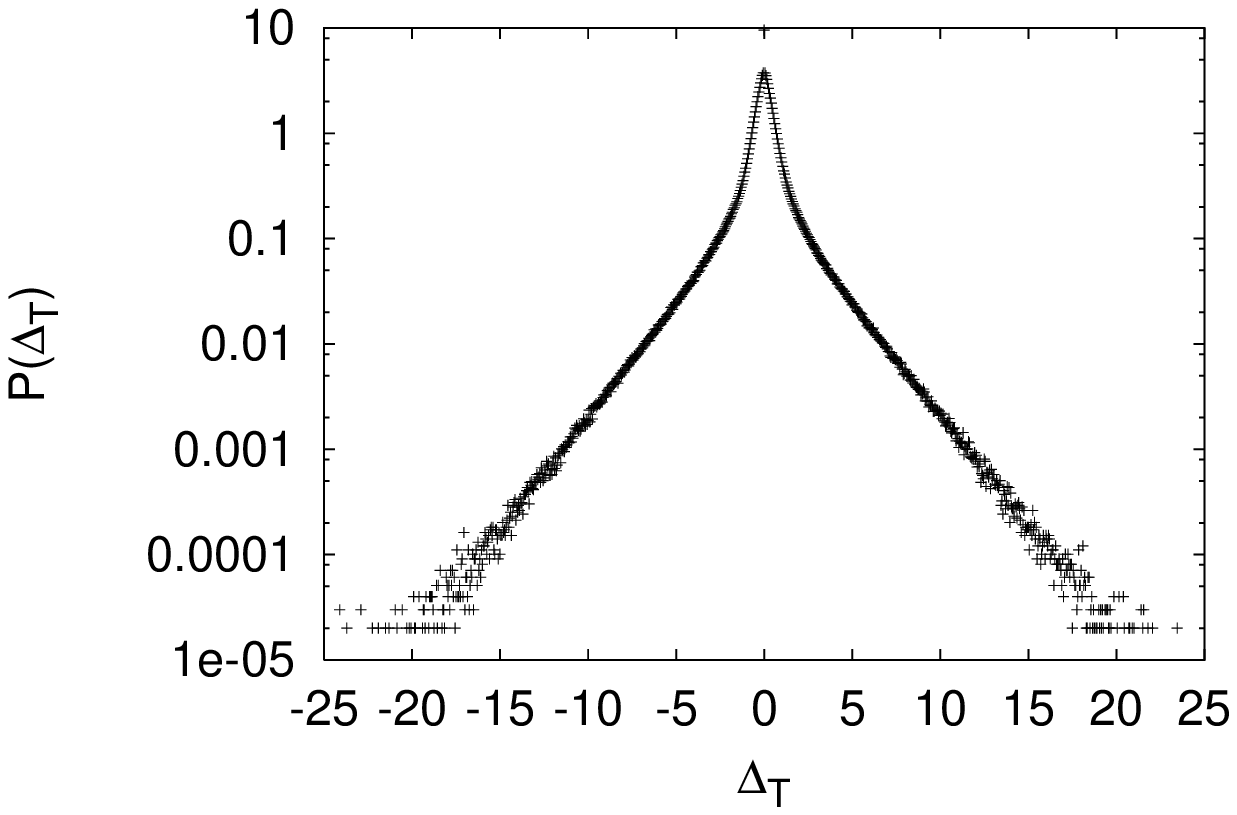}
\caption{The same as Fig. 1, however with interactions $I_1$ (see
text) and $x=1$ (i.e. investors look only at the old price
$K(t)$). Note the spikes in the time dependence of the price marking
the significant enhancement of price fluctuations that lead to the
truncated Levy-distribution of the price changes.
\label{fig4}
}
\end{figure}
\vfill
\eject
\begin{figure}

\mbox{
\epsfxsize=8cm\epsfbox{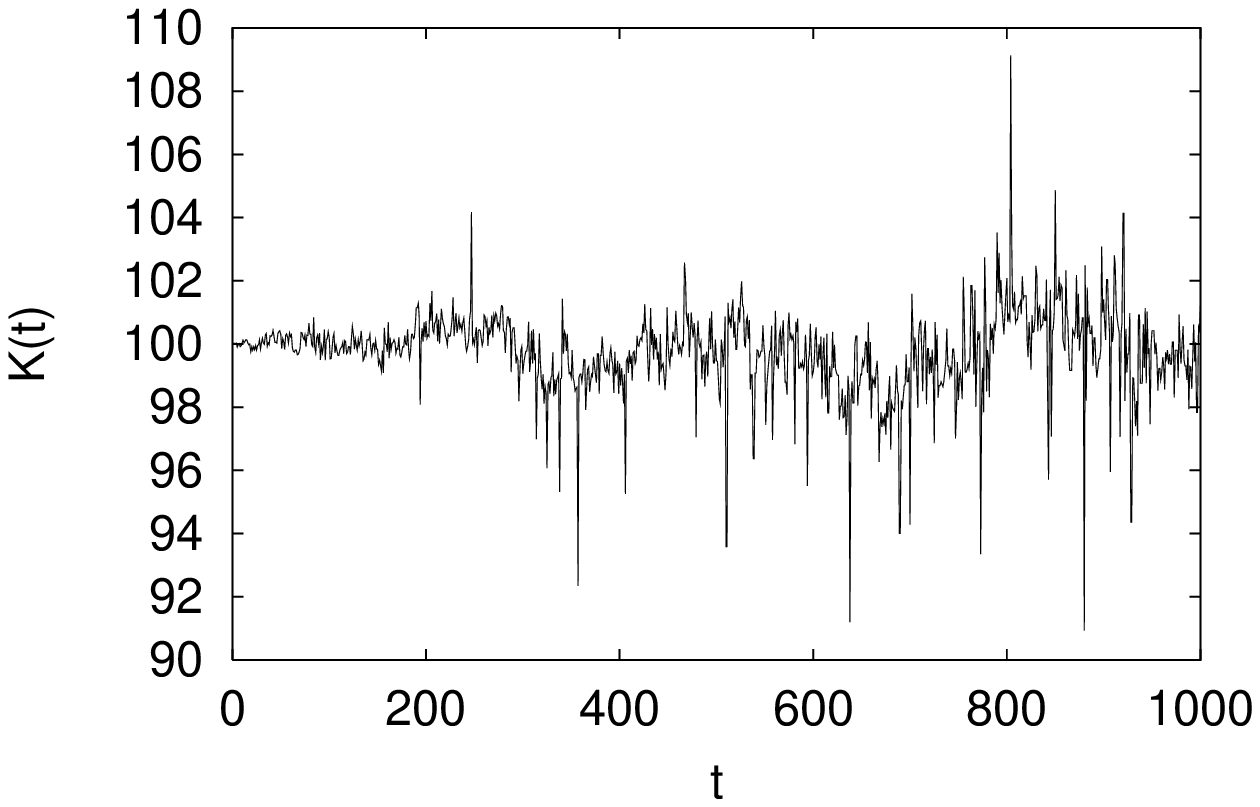}
\epsfxsize=8cm\epsfbox{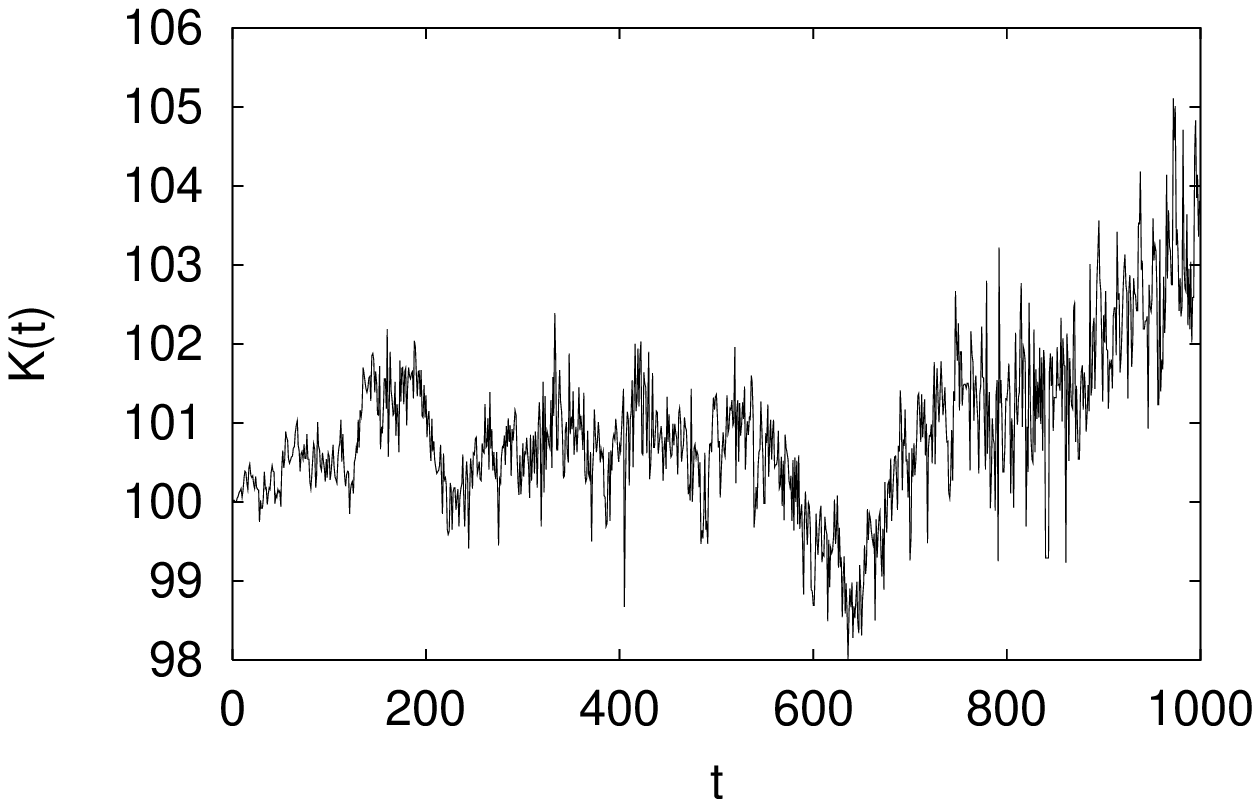}
}
\mbox{
\epsfxsize=8cm\epsfbox{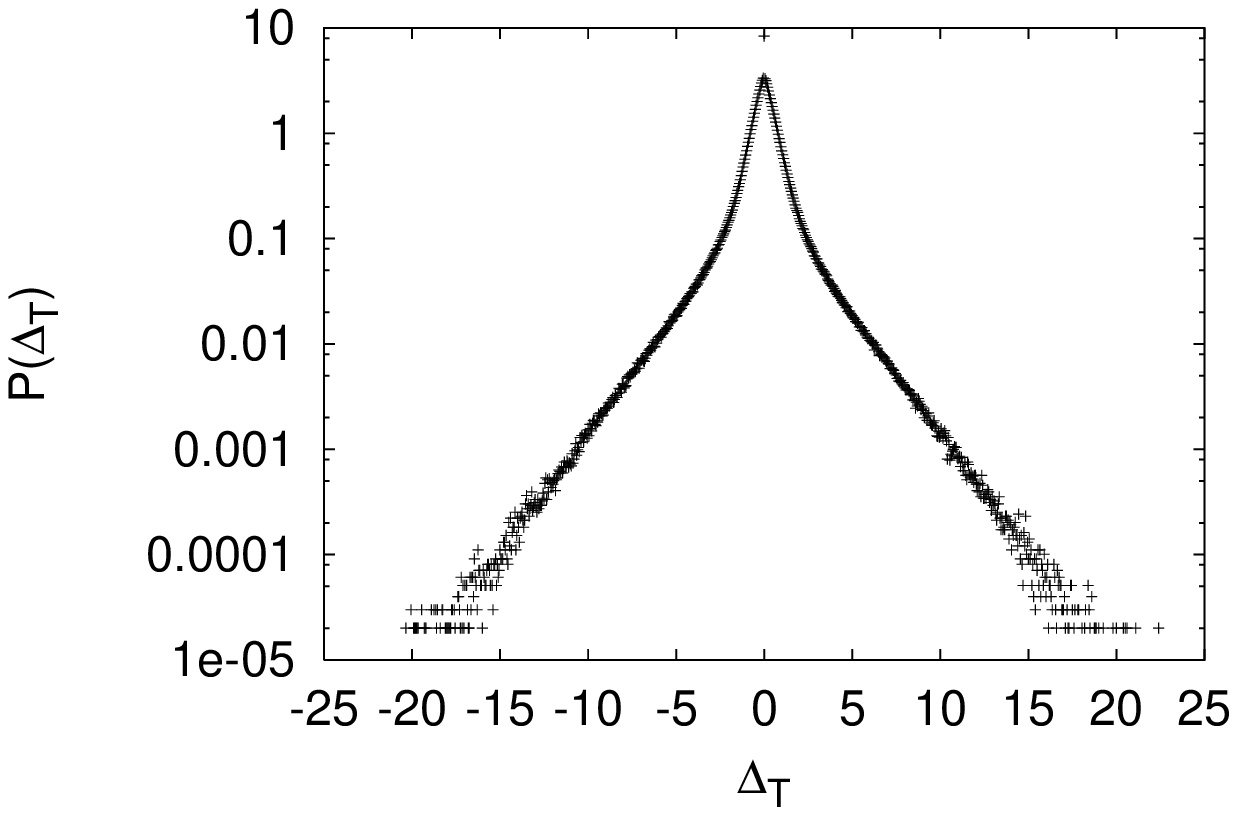}
\epsfxsize=8cm\epsfbox{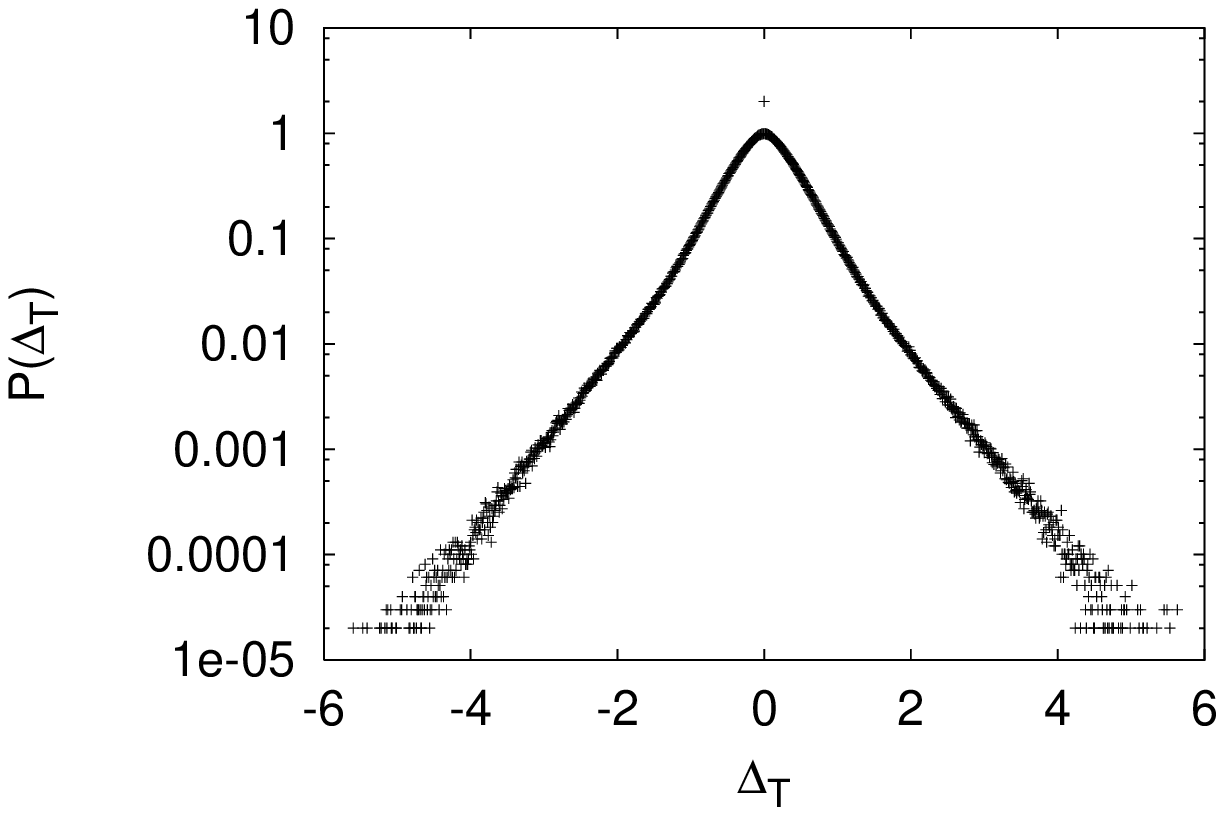}
}
\caption{The price fluctuations $K(t)$ (top) and the price difference
distribution $P(\Delta_1)$ (bottom) of the model {\it with}
interactions of the investors $I_2$ (left) and $I_3$ (right).  The
delta peak at $\Delta_1=0$ comes from the events were no trade took
place.
\label{fig5}
}
\end{figure}
\vfill
\eject

\end{document}